# Towards Best Practices for Open Datasets for LLM Training

*Proceedings from the Dataset Convening[1]*

**Stefan Baack (1), Stella Biderman (1), Kasia Odrozek (1), Aviya Skowron (1), Ayah Bdeir (2), Jillian Bommarito (2), Jennifer Ding (2), Maximilian Gahntz (2), Paul Keller (2), Pierre-Carl Langlais (2), Greg Lindahl (2), Sebastian Majstorovic (2), Nik Marda (2), Guilherme Penedo (2) Maarten Van Segbroeck (2), Jennifer Wang (2), Leandro von Werra (2), Mitchell Baker (3), Julie Belião (3), Kasia Chmielinski (3) Marzieh Fadaee (3), Lisa Gutermuth (3), Hynek Kydlíček (3), Greg Leppert (3), EM Lewis-Jong (3), Solana Larsen (3), Shayne Longpre (3), Angela Oduor Lungati (3), Cullen Miller (3), Victor Miller (3), Max Ryabinin (3), Kathleen Siminyu (3), Andrew Strait (3), Mark Surman (3), Anna Tumadóttir (3),  Maurice Weber (3), Rebecca Weiss (3), Lee White (3), Thomas Wolf (3)**

(1) Project leads (2) Top contributors (3) Contributors (all alphabetized within tier)

## Abstract

Many AI companies are training their large language models (LLMs) on data without the permission of the copyright owners. The permissibility of doing so varies by jurisdiction: in countries like the EU and Japan, this is allowed under certain restrictions, while in the United States, the legal landscape is more ambiguous. Regardless of the legal status, concerns from creative producers have led to several high-profile copyright lawsuits, and the threat of litigation is commonly cited as a reason for the recent trend towards minimizing the information shared about training datasets by both corporate and public interest actors. This trend in limiting data information causes harm by hindering transparency, accountability, and innovation in the broader ecosystem by denying researchers, auditors, and impacted individuals access to the information needed to understand AI models.

While this could be mitigated by training language models on open access and public domain data, at the time of writing, there are no such models (trained at a meaningful scale) due to the substantial technical and sociological challenges in assembling the necessary corpus. These challenges include incomplete and unreliable metadata, the cost and complexity of digitizing physical records, and the diverse set of legal and technical skills required to ensure relevance and responsibility in a quickly changing landscape. Building towards a future where AI systems can be trained on openly licensed data that is responsibly curated and governed requires collaboration across legal, technical, and policy domains, along with investments in metadata standards, digitization, and fostering a culture of openness.

On June 11, 2024, Mozilla and EleutherAI convened 30 scholars and practitioners to create normative principles and technical best practices for creating openly licensed LLM training datasets. Based on that convening, this paper outlines the challenges of navigating the production of open datasets and provides practical recommendations for sourcing, processing, governing, and releasing of these datasets. It seeks to be the foundation for shared practices and long-term goals within the emerging community around LLM data and to bring the community closer to making this technology [truly open and trustworthy](#).

---

[1] The Dataset Convening was inspired by the format of the Columbia Convening on Openness in AI, a collaboration between Columbia University's IGP and Mozilla, and held in February 2024 in New York. More details here and full publication here.

## 1. Introduction

Today's AI systems depend on the data used to train large language models (LLMs), making dataset transparency critical for accountability and innovation. While AI companies used to be more open about their training data, as seen with Google's [T5](#) or Meta's [LLaMA 1](#), little is known about the datasets behind the most popular models from companies like OpenAI, Anthropic, Google, and Meta today. Over the past year, AI companies have faced heavy criticism, particularly from literary and creative communities, over perceived exploitative data practices, leading to [multiple](#) [high-profile copyright lawsuits](#). Regardless of these lawsuits' legal merits, they have become one of several factors discouraging companies from disclosing their data sources and governance practices.

Meanwhile, there is an ecosystem of open LLM developers — startups, researchers, and nonprofit organizations — who are increasing transparency in AI training data and promoting open access to those datasets. In June 2024, Mozilla and EleutherAI convened 30 scholars and practitioners from prominent open-source AI startups, nonprofit AI labs, and civil society organizations working on open access and openly licensed datasets for the [Dataset Convening](#). After analyzing case studies from three of the leading open datasets (EleutherAI's forthcoming Common Pile, Pleias' [Common Corpus](#) and [YouTube-Commons](#)), the group met to discuss the most pressing challenges and opportunities around creating open-access and openly licensed LLM training datasets. The group identified seven principles to guide the creation of these datasets:

1. **Foster a competitive LLM ecosystem**
2. **Enable accountability and transparency through reproducibility**
3. **Minimize harms and enable preference signals**
4. **Support and improve diversity**
5. **Strive for reciprocity**
6. **Work with other like-minded actors in this space**
7. **Preserve data for the long term**

The group also identified the challenges and pitfalls facing organizations seeking to build these datasets. This paper captures those insights, including emerging normative and technical recommendations for the community. The Dataset Convening was part of a longer series of events co-hosted by Mozilla, inspired by (and including) the [Columbia Convening (February 2024)](#) which developed a framework for openness in AI, and continuing with an upcoming Columbia Convening on AI Openness and Safety (November 2024).

The authors of this paper view openly available and openly licensed datasets as important because:

1. They can serve the public good by enabling developers to build upon others' work efficiently.
2. They can themselves be public goods that everyone can use without reducing their availability to others.
3. They create the right incentives for volunteer-driven or [data donation-based processes](#).
4. They enable scrutiny, supporting AI accountability by facilitating [new lines of inquiry for auditors and researchers](#), particularly in areas requiring manipulation of model training data.

Many of the challenges faced by open dataset builders today bear a resemblance to those encountered in the early days of open source software, such as data quality, standardization, and sustainability, especially because the ecosystem — just as back then — often relies on community contributions and volunteers. With open source software, it was common artifacts such as software projects, licenses, and documents that united the community and provided shared understanding and language.

This paper is intended to help create a similar shared reference point for the emerging open LLM ecosystem and help the community orient in a common direction when it comes to open datasets. It complements efforts like the [Data Ethics Canvas](#), [The Model Openness Framework](#), [Foundation Model Development Cheatsheet](#), [Data Governance in the Age of Large-Scale Data-Driven Language Technology](#), [Building a Better Future with Data and AI](#) or the [Open Source AI Definition](#).

While the paper is not intended to be a comprehensive analysis of all considerations around open datasets for LLM training, it does aim to capture the ambitions and consensus that is starting to emerge from key scholars and practitioners working in this field, and to identify prominent challenges and open questions going forward.

## 2. Terminology

There are varying definitions of dataset openness in the AI community. In this paper, we focus on three types of openness in AI data:

- **Openly licensed dataset**[1]: A dataset and its components can be freely used, modified, and shared by anyone for any purpose (following the Open Knowledge Foundation's Open Definition for data and content).
- **Downloadable/open-access dataset:** Dataset is available to freely download, with no claim about license compliance.
- **Replicable dataset:** Data sources and processing steps are disclosed, such that an independent party can produce a substantially similar (albeit not identical) dataset, or what the Open Source AI Definition calls "a substantially equivalent system." This assumes data sources are widely accessible (e.g., not internal data or data accessed through private agreements).

## Tiers of dataset openness

|  | Tier 1 | Tier 2 | Tier 3 (Fully open) |
|---|---|---|---|
| Legal status: enables the dataset user to reuse, share, modify. |  |  | Openly licensed |
| Data availability: enables the dataset user to download. |  | Open access | Open access |
| Sufficient Documentation: enables the dataset user to understand and reproduce. | Replicable | Replicable | Replicable |
| Examples | CR4 mC4 | Pile Dolma FineWeb | Common Pile (forthcoming) Common Corpus |

Figure 1. Tiers of openness of datasets for LLM training.

An important distinction here is between the licensing of the dataset and the licensing of the constituent parts. In some jurisdictions, the act of collecting, processing, and arranging existing data grants intellectual property interests *in the arrangement.*[2] However, it does **not** grant the compiler the right to change the licensing of the underlying data. This is an essential distinction, but one that is not widely known, and can cause confusion and incompatibility when the license of the dataset (based on rights in the arrangement of the dataset as a whole) is substantially more permissive than the licensing of the individual components. Throughout this document, when we discuss licensing, we are referring to **both** of these types of IP being suitably licensed, as the construction of the actual input to a machine learning model involves both.

---

[1] We use the term "dataset" to include all of a dataset's individual components. According to the Linux Foundation, "data can be any form or combination of media, whether text, code, images, videos, audio, 3D objects, URIs, and any other data used for training, validation, and testing purposes. Datasets also include any metadata, from annotation data, such as labels, bounding boxes, and key points, to attribution, bitrates, resolution, and other metadata that may be relevant to a dataset used in the model development process." White, Haddad, Osborne, Abdelmonsef: "The Model Openness Framework: Promoting Completeness and Openness for Reproducibility, Transparency, and Usability in Artificial Intelligence", *arxiv:2403.13784v3*, June 2024.

Similarly, some datasets like LAION do not technically contain the core objects of interest for training machine learning models. In LAION's case, it contains pairs of (caption, url-of-image) and *not* the images themselves. Another example is C4, as originally released by Google, which contains the code and other information required to *construct* the dataset and not the dataset itself. Again, as the focus of this paper is on the production of datasets for training machine learning models, we are interested in the licensing of all relevant materials and objects, even if they are not technically part of the officially distributed dataset.

**While these may seem like minor technicalities, we have observed substantial confusion about the licensing of machine learning datasets due to these points.** For example, many people believe that AI2's Dolma dataset is openly licensed due to the ODC-by dataset license even though the constituent data is not openly licensed.

Openness, ethics, and legal compliance are notions that simultaneously intersect and compete with one another.

- **On openness**: Open access is a set of publishing practices that enable free online access to research outputs, such as scholarly papers and experiment data. Open source refers to a set of standardized software licenses, and a broader culture of co-creation and sharing of code that has become the foundation of modern technology.
- **On compliance:** There are several areas of law that are relevant to dataset construction, such as intellectual property, data privacy regulation, and contract law. Compliance is about putting these legal requirements into practice. Due to uncertainty about how to apply existing law to new technology, AI actors today necessarily assume a level of legal risk. Because each actor has different risk tolerance, legal decisions vary greatly.
- **On ethics**: It is crucial to remember that an action can be legally compliant without being ethical. Ethics concerns itself with wider questions of, for example: justice, equality, distribution of resources, redress of harms. A project can align with a set of ethical values while falling short of legal compliance. All three of these areas of focus are evolving rapidly in response to new AI developments.

---

[2] E.g., [EU sui generis database right](#).

Dataset builders need to identify legal, ethical, and practical goals and consider how they interact. Sometimes these goals will be in conflict with one another, such as when implementing a rolling individual opt-out for content creators contradicts the idea of working with a verifiably identical dataset over time.

This document outlines the challenges of navigating both the definitional and executional aspects of open datasets, and provides practical recommendations for sourcing, processing, governing, and releasing these datasets.

## 3. Challenges and Guiding Principles

Open dataset builders face a myriad of challenges including legal, technical, operational and more. Shared principles help articulate common goals and strategies and navigate resolutions as the landscape changes.

### 3.1. Challenges

Building and releasing an open access dataset can be a complex technological and legal problem that requires collaboration and expertise. The collection, identification, and validation of a large-scale openly licensed dataset can require substantial amounts of manual work, consultations with legal experts, and technological skill — despite improvements based on innovations in language and image modeling.

Challenges of open datasets include:

- **Laws vary across jurisdictions and time**. Developers of LLMs span the globe and speak numerous languages, and copyright law varies by jurisdiction. Determining whether a particular document is in the public domain can require country-specific analyses, and it can require many lawyers to vet work published in multiple countries. Laws can also change over time, resulting in a web of more complicated interlocking requirements.

- **Relevant metadata is incomplete**. What constitutes a "work" under copyright law does not necessarily correspond to one dataset document, electronic file, or HTML tag. This can significantly limit the usefulness of existing license information. For example, when filtering the Common Crawl, it is easy to determine that a website links to a CC-BY 4.0 license, and therefore contains some sort of CC-BY 4.0 statement. However, there is currently no automated way of determining which asset on the website that license covers. This leads to false positives if, for example, someone uses a third-party CC-BY 4.0 photograph as part of their non-CC article. Metadata challenges also apply to determining if work is in the public domain.[3] Unfortunately, there is no official database for such information: third parties have largely converted the original US copyright renewal forms into digital text, but matching renewal submissions to original applications can be challenging due to data quality issues and differences in forms. While the U.S. Copyright Office does assign a numerical ID to copyrighted works, that ID is not necessarily unique because the numbering system has been changed multiple times in the past century. Somewhere around 480,000 books published between 1929 and 1989 are estimated by the New York Public Library to be in the public domain because their copyright status was not renewed, but the specific titles of these books have yet to be identified.

- **Just because a document is in the public domain, it does not mean one can get a copy of it.** Many books in the public domain have never been digitized. For such books that have been digitized, it can often be challenging to get access to that book. A large portion of the world's digitized books were scanned by Google in partnership with libraries as part of the Google Books project. While some access to these digitized books is possible through the Google Books platform, it is not possible to gain unrestricted bulk access to all books that even Google thinks are in the public domain. Even when access is granted, organizations often need to sign agreements that limit their ability to use the content. This mirrors the struggle to gain unlimited public access to copies of texts or photos of artwork that are physically owned by cultural heritage institutions, but whose intellectual property has passed into the public domain.

- **Managing legal risks in a volunteer-driven, decentralized group of contributors**. Many open source projects are organized in a collaborative but unstructured fashion: volunteers from around the world contribute what they want to work on and decisions about the project's direction and standards are made with no formal process. Furthermore, many open source projects do not have a legal entity that is ultimately responsible for the project's outputs. This is challenging in situations when there is a serious risk of litigation, which often require top-down decisions guided by dedicated attorneys bound to attorney-client privilege. A legal entity that has ownership of, and responsibility for, the project can also limit the individual liability of contributors.

- **Preventing data enclosure without further cementing the market advantage of incumbents**. As shown by the Google Books example, even public domain data can remain unusable for open dataset construction once digitized. This suggests there is a need for serious public support for open data commons — we cannot expect data will be available simply because they are not protected by copyright. Creating this support is in itself a policy challenge, but it must happen in parallel with investment into new digital infrastructure for communicating opt-outs (section 3.2).

---

[3] For example, works published in the U.S. before March 1, 1989, are also in the public domain if the copyright was not registered with the U.S. Copyright Office within five years of the date of original publication.

## 3.2. Guiding Principles

This paper proposes the following guiding principles to help dataset builders in their work to create open datasets for LLM training:

1. **Foster a competitive LLM ecosystem**: A handful of tech companies should not have outsized control over LLM research and development. To avoid this, builders should provide competitive alternatives and base layers for other developers to build on top of. Creating transparent open datasets, which can be audited more widely, can help mitigate the legal risks for the training and application of open source AI models and help such models be competitive with the closed AI models. This promotes competition, as smaller actors are often concerned about legal exposure.
2. **Enable accountability and transparency through reproducibility:** LLM training datasets need to have more transparent production pipelines. Developers should strive to provide reasoning for all steps in the data collection and filtering process, as well as access to tooling and source code for others to replicate their process. This is vital for auditing the model development process and increasing accountability for the model developers. It is also fundamental for research as one cannot improve upon best processing setups if they are not known.[4]
3. **Minimize harms and enable preference signals**: Standards are needed throughout the data production process. The goal should not be creating the "perfect" dataset, but rather, developing interoperable standards for data governance in order to provide easy ways for data subjects and rights holders to declare their preferences before model training (e.g., at the moment of data collection) and to report issues thereafter. Recognizing that people or organizations might want to opt out, dataset builders should have a plan for how to remove content from the dataset. However, it is important to acknowledge that such removal processes could limit both reproducibility and transparency, if not well documented, as well as the competitiveness of open datasets and data availability for research when faced with massive opt-outs on the open web.[5] See section 3.4 on data governance for more discussion around this issue.
4. **Support and improve diversity**: The quality and coverage of training data in different languages and representing a diversity of cultures often varies greatly. Voice and text datasets powering AI dramatically under-represent 99%+ of global languages, variants and dialects, as well as Black, Indigenous and People of Color and gender-diverse communities. In order to support LLMs that can be the foundation for an open-ended set of applications around the world, there must be a diversity of languages and viewpoints represented in datasets.
5. **Strive for reciprocity**: Data collection should be mutually beneficial and reciprocal. Today, data subjects, data contributors, organizations and rights holders do not get direct benefits (monetary or otherwise) out of their data being included in LLM training datasets. A better process would go beyond simplistic yes/no mechanisms like robots.txt and find ways to empower communities, creators, and others with a legitimate interest regarding the data, preventing their exploitation. It also means finding ways to convince and find reciprocity with larger institutions, who often hold valuable data but can be incentivized to make data licensing agreements with AI companies, to make their data more open.
6. **Work with other like-minded actors in this space**: Organizations like Wikipedia and Creative Commons, open science projects, open data initiatives, libraries, and more have relevant expertise and can help address issues with LLM training datasets.
7. **Preserve data for the long term:** Training datasets for AI should ensure that the data is interoperable and the information contained in the dataset will be preserved and remain accessible in the long term.

---

[4] See also Warso, Gahntz, Keller: "Sufficiently Detailed? A proposal for implementing the AI Act's training data transparency requirement for GPAI", *Open Future and Mozilla*, June 2024.
[5] See also Longpre, Mahari, Lee et al.: "Consent in Crisis: The Rapid Decline of the AI Data Commons", Data Provenance Initiative, July 2024.

## 4. Best Practices

This section provides recommendations for best practices to advance the aforementioned principles. These are non-exhaustive and intend to combine the insights from EleutherAI's work on the Common Pile (c-pile) ([Appendix A](#)), Pleias' work on Common Corpus and YouTube-Commons ([Appendix B](#)), and the knowledge of the participants at Mozilla and EleutherAI's dataset convening.

### 4.1. Encoding preferences in metadata

Finding content that is openly licensed or public domain across jurisdictions is difficult and often requires "artisanal" manual labor. While this is not an inherent part of the dataset production pipeline, we recognize the need for developing standards that make the data processing pipeline (and the web itself) more likely to provide accurate and complete metadata, particularly in complex online environments that may have multiple licenses and hierarchical terms.

The advantage of implementing machine-readable preference signals and preserving metadata through processing is that it enables data governance downstream. It serves as a necessary building block for many existing and proposed mechanisms, such as copyright holder opt-outs. Usable metadata is the crucial first step to realizing many of the goals outlined in this document.

4.1.1. Emerging Best Practices

- **Identify and preserve relevant metadata.** Examples include URLs and licenses associated with given content. For interoperability, we recommend using existing tools such as [SPDX license identifiers](#). This is to enable future compatibility with preference signaling tools that are currently being developed and adopted.
- **Develop and adopt machine-readable standards for content identification and preference signals.** This is necessary for building data governance and consent infrastructure for the internet. Due to EU legislation, "general-purpose AI" (or "foundation model") model developers need to adopt, and ideally converge on, [machine-readable opt-out standards](#) by August 2025 in order to comply with the EU AI Act, which references the EU Copyright Directive's Text and Data Mining (TDM) exception. In addition, much of the public reaction to generative AI emphasizes the need for better data governance in order to demonstrate good faith engagement between technologists and creatives.

4.1.2. Examples of emerging protocols

As the landscape of opt-out mechanisms develops, it is important to distinguish between (A) mechanisms that work on the network blocking level (e.g., [Cloudflare's "One-click to block all AI bots"](#)) which effectively prevent scraping of websites and need to be weighed against the mentioned problem of shrinking the open web, and (B) tools that focus on preference signals that rely on voluntary compliance by the crawling party, much like the decades-old [robots.txt](#) and the recent developments focusing on more nuanced and granular preference signaling at the level of data rights holders. Examples of these include:

- **International Standard Content Code (ISCC)** is an ISO standard for [creating unique digital identifiers](#) that work regardless of medium (text, audio, image, video).
- **Spawning** is building [solutions for machine-readable opt-out methods](#), making it possible for copyright holders to express preferences, and offers an API for AI developers.
- **Creative Commons** is working on [preference signaling for AI data](#) that allows for more granular control than simply fully allowing or disallowing AI use cases.
- **BigCode opt-out process for The Stack** is [a manual option](#) to exclude or remove repositories from the Stack dataset that relies on Github accounts to verify the user's identity and reasonably validate their data rights.

## 4.2. Data Sourcing

The most capable LLMs are typically trained on large amounts of diverse and high-quality data. Yet what exactly this means is often ambiguous. Those compiling and providing open and responsible data sources have a lot of influence over this data ecosystem.

4.2.1 Emerging best practices

- **Prioritize community resources:** Where possible, rely on community-driven tools and resources for identifying and collecting data, and openly make available (custom) tools developed in the process.

- **Provide useful documentation:** For proper data documentation and to help with auditing datasets, make it easy to fully replicate the data sourcing process. This involves describing why sources were chosen, how data was acquired from them, and sharing the source code of tools that were used in the process. When synthetic data is used, provide the full tooling for how this data was generated, including information on the source of the seed data, the prompts and model(s) that were used to generate synthetic data, as well as the implementation of privacy-preserving methods like differential privacy to protect individual data points, and anonymization of data to remove PII when fine-tuning models to generate synthetic data.

- **Follow and record preference signals:** For each data point, record the associated permissions and the metadata needed to determine them (such as url, crawl date, http headers and html metadata), if available, as well as methods used to determine them. This refers to respecting signals such as robots.txt and licenses associated with code repositories and content, as well as any future implementation of data governance signals.

- **Increase diversity and involve local communities to identify relevant data sources:** Language and regional coverage should not only be measured in quantity; the quality and context of the sources matter. For general purpose datasets, employ a mix of data sources to capture a broad spectrum of content, and ensure each source is evaluated for its specific benefits and challenges related to diversity and quality.

- **Do not rely heavily on automated translations to include more languages:** Many LLM training datasets are primarily in English. However, trying to counter this lack of representation with automated translations often backfires, as their quality is bad and ignores culturally specific aspects, especially for minoritized languages and low-resource languages.

- **Share advancements to foster reciprocity and give back:** Make any open content sourcing enhancements — such as PDF parsing or Optical Character Recognition — available to the commons ecosystem and give back to the right holders and those who enabled your work. For example, the team behind the Common pile sent transcripts of all the videos they used to the YouTube channels and the work they did on identifying creative commons licensed subsets of Common Crawl is being gifted to Common Crawl for them to use to improve the metadata in Common Crawl.

- **Use synthetic data with care:** Synthetic data is no silver bullet; it comes with its own biases and problems, and should be used with care just like any other data. Use quality metrics to ensure synthetic data is accurate, consistent, and representative of the original dataset. Regularly inspect the synthetic data to maintain ethics, privacy, and quality standards.

- **Do not use openly licensed data without regard for its quality or fitness for purpose:** Do not try to inflate the size of training datasets with low quality data — the quality of the data matters.

- **Do not capture highly sensitive data:** Take steps upfront to avoid collecting sensitive data like phone numbers or health information.

4.2.2. Examples of emerging data sourcing practices

- **Common Pile:** A work-in-progress dataset that aims to be a fully transparent dataset for training LLMs composed exclusively of public domain and open access data.

- [Common Corpus](...) (first release) is a multilingual, downloadable, and reproducible public domain dataset. Pleias later [expanded Common Corpus](...) to include permissibly licenced text, using the updated dataset to train and fine-tune models.

- [Dolma](...): An extensively documented downloadable dataset that explicitly outlines design principles, details about its construction, and a summary of its contents.

- [FineWeb](...) **and** [RefinedWeb](...): Two downloadable datasets that fully rely on Common Crawl, a massive archive of web crawl data, and both provide very detailed documentation on what parts of the archive were included.

- [Mozilla Common Voice:](...) A public participation, openly licensed dataset in 120+ languages, created and curated by volunteers around the world. All text and audio is CC0.

- **Aya Dataset**: An openly licensed human-annotated multilingual dataset curated by an open-science community released with comprehensive documentation and guidelines and code of the annotation tool. Released under Apache-2.0 license.

- **Big Code** is a collaborative project supported by Hugging Face and ServiceNow. All datasets, models, and experiments are developed through open collaboration and released with permissive licenses back to the community.

- **PD12M** by Spawning is the largest public domain image-text dataset to date. Compiled to resolve the data quality issues that arise in web-scraped training data: the presence of copyrighted material, low quality images and captions, violent, hateful, or NSFW content, PII, decaying dataset quality via broken links, etc. Released under CDLA-Permissive 2.0 license.

- **Data Provenance Explorer**: a catalog from a large-scale audit of AI datasets that filters for text fine-tuning datasets with permissive and open-source licenses.

- **Project Gutenberg** is a dataset of literature classified as public domain in the US (poetry, short stories, drama, cookbooks and more, mainly from Western origin) comprising about 70,000 books. Released under The Project Gutenberg Licence.

- **KL3M:** A project providing a replicable pipeline from data collection to model training that is certified under the Fairly Trained L-Certification standards for training data. The KL3M dataset only contains sources with explicit legal authority or for which consent has been obtained from rightsholders, reducing uncertainty related to other "open" sources (e.g., CC-BY-SA/GFDL content like Wikipedia, which is still subject to legal action by each individual Wiki contributor under copyleft terms).

**4.3. Data Processing**

Careful attention to data processing and cleaning are crucial for ensuring datasets comply with licenses and are technically robust. The approach varies significantly based on the data source; for example, Common Crawl and web data requires different handling than targeted data collections. Understanding exactly how each source was preprocessed is crucial.

4.3.1. Emerging best practices

- **Clearly and explicitly state the values and desired properties that shaped the way data was filtered or annotated**: "High-quality data" is a term that is used a lot, but it is not a properly defined concept. It needs to be defined in relation to the dataset that is being curated. That also means acknowledging that not all potential harms and risks can be mitigated directly via interventions in the dataset, as it may be used in many different contexts. The filtering and processing goals will also vary depending on the intended use of the AI system, such as whether it is designed for open generation or constrained to specific tasks, or whether end users have more or less awareness about the system's potential issues. The values in data processing and the definition of "high-quality data" thus needs to be tailored to the specific application and user base.

- **Strive for reproducibility**: Provide documentation that outlines the rationale for all steps in the filtering process. Share tools and code used to filter the data. If data workers were employed, describe the recruitment process, working conditions and guidelines they had to follow. This makes auditing easier, helps to spread documentation best practices, and can help support better employment conditions for data workers.

- **Attempt to identify content that does not align with stated values**: This includes not only harmful content, but also content that promotes harmful outcomes in downstream applications. Depending on the context, this data can either be filtered out, or annotated for data provenance reasons allowing downstream users to decide how to use it based on their specific use cases.

- **Consider potential unintended consequences of your filtering methods**: Your filtering introduces its own biases, and can cause harms if used without care. For example, simplistic word blocklist based filtering techniques might filter out non-toxic content, such as medical research articles discussing anatomy.

- **Uphold existing standards**: At minimum, follow established transparency best practices like [data sheets](#) or [data cards](#).

4.3.2. Examples of data processing practices

- [FineWeb](#) **and** [Dolma](#)**:** These projects demonstrate the effective removal of personal identifiers like email addresses and phone numbers. [FineWeb Edu Classifier](#) is an example of a reproducible tool for data cleaning. The full [FineWeb pipeline](#) is made available on Github.

- **Hugging Face**: As a platform, Hugging Face provides guides for the creation of datasheets and [data cards](#), as well as practical information and demos for AI developers on their [Ethics & Society](#) space.

- [Toxicity classifier](#)**:** open-source [pipeline for open-data toxicity filtering](#) by Pleias.

**4.4. Data Governance/Release**

Data governance refers to the rules and processes for how data is collected, accessed, controlled, used or shared. Ideally, training data for LLMs should be governed in ways that are inclusive, empowering, and mitigate harms.

4.4.1. Emerging best practices

- **Tailor data governance mechanisms to your data subjects and use case:** Not every dataset needs to be open access. For example, workshop participants described projects for which they developed dataset gating mechanisms at the request of data subjects. Open access datasets can coexist with more access-restricted datasets because they frequently concern different types of data. Common Pile is focused on texts that are public domain or hosted on open access repositories; on the other hand, datasets with more targeted access privileges tend to be smaller in scale and more personal, such as collections of recordings made by living vocalists, or are created for the benefit of particular communities (e.g., work by [Ushahidi](#) and [First Languages AI Reality](#)). However, all approaches need usable metadata and accessible documentation that communicates this to users, thus emphasizing the importance of the data governance building blocks outlined in sections 4.1 and 4.2.

- **Work with affected communities:** Communities and organizations impacted by AI dataset development should be meaningfully engaged with as stakeholders (e.g., data trusts for language communities, labor unions representing writers or artists).

- **Post-release removal**: Create models of redress and removal from a dataset if an issue is spotted. For example, from the start provide mechanisms for people to request removal of their data, and encourage downstream users of the dataset to only use the updated version. Note that this is only possible if sufficient content identifiers are available. It is important to acknowledge the tension between opt-outs and the competitiveness of open datasets. Current mechanisms [focus](#) on who is allowed to crawl a website rather than how its data is used, which leads many website owners to completely block non-profit archives like Common Crawl that are used for non-commercial purposes by researchers and non-profits. Improving this, alongside more reliable opt-out mechanisms in existing datasets, might also help counter the decline of openly available content on the internet.

- **Strive for accessible transparency**: Make it easy for people without a technical background to check if their data is in a dataset.

- **Strive for socially beneficial uses**: Think about ways to encourage positive uses of the dataset, such as through promoting good use cases and specifying intended uses in the data card.

- **Control the versioning:** Dataset builders often release their datasets on multiple platforms like HuggingFace and their own websites. Consider where you release your datasets and how it influences your ability to control, maintain and update them consistently across platforms.

4.4.2. Examples of emerging frameworks for governance and metadata documentation

- **Community-based data trusts or trusted data intermediaries:** legal structures for managing and collectivizing rights and preferences of dataset contributors to enable collective governance of the dataset (e.g. a [legal framework for the UK Choral AI Dataset](#)).

- **BigCode opt-out process for The Stack** is [a manual option](#) to exclude or remove repositories from the Stack dataset that relies on Github accounts to verify the user's identity and reasonably validate their data rights. BigCode also provides a [tool](#) to verify if one's data is in the dataset, as well as a [Project Governance Card](#) that serves as an overview of the different mechanisms and areas of governance in the project.

- [Spawning](#) is offering opt-outs from existing datasets via an API integrated in Hugging Face and [community-driven dataset governance mechanisms](#) for managing datasets, reducing harm and supporting reproducibility over time.

- **Croissant** is a [metadata documentation standard](#) for ML ready datasets by the ML Commons community that has been adopted by platforms such as Hugging Face and Kaggle.

- [Data Provenance Standards](#) by Data & Trust Alliance, including companies such as IBM and Deloitte, are cross-industry metadata standards aiming to bring transparency to the origin and legality of datasets used for both traditional data and AI applications.

**4.5. Terms of Use**

If terms of use are provided, they need to be clear and aimed at ensuring that the data is used responsibly, respecting the rights of data subjects, and aligning with global integrity-based standards. Terms of use should only be provided if there are clear pathways to enforce them.

4.5.1 Emerging best practices

- **Standardization and modularization**: Many datasets have their own terms of use and categories of harm are different. A good step forward is developing systems for creating modular terms of use that are technically recognizable and easily adaptable.

- **Accessibility**: Design terms of use that are clear and centered around user needs and understanding by making terms of use accessible and understandable to non-legal professionals, reducing barriers to compliance.

- **Do not impose restrictive terms on public domain data**: It is equally important to ensure that data that now belongs to the public are made maximally useful, and avoid adding restrictions to public domain works (e.g. via the misuse of Creative Commons licenses).

4.5.2. Examples of emerging frameworks

- [Responsible AI Licenses (RAIL)](#) aim to empower developers to limit particular uses of their AI technology in order to mitigate irresponsible and harmful applications via [behavioral use licensing](#). Currently, RAIL has been adopted by about 40,000 software and model repositories and their creators advocate for ["standardized customization"](#) of use to avoid diluting of their impact while allowing for nuanced use. About [8000 datasets with RAIL licenses](#) can be explored on Hugging Face for instance.

## 5. Recommendations for the tech and policy community

Throughout this document, we've outlined both short- and long-term recommendations for technological and policy improvements, examining how decisions around datasets will ultimately influence the long-term openness of the Web. This section synthesizes key challenges and suggests actionable steps that stakeholders in the AI ecosystem might consider moving forward.

**5.1. Open data availability**

The process of identifying licensing status and metadata across jurisdictions can be overwhelming, causing valuable data to remain inaccessible. A lot of open data is locked in inaccessible or gated repositories or formats and the one that is accessible is often unstructured. Many companies or institutions don't even know that and how they could release their data into the open. Finally, massive opt-outs of AI crawlers threaten to significantly reduce open data availability.

**Policy interventions**

- Simplifying the identification of public domain data internationally would reduce complexity and labor, enabling better use of open data and datasets. Institutions like the EU and public libraries could play a pivotal role by certifying public domain content, streamlining data availability.

- Requiring certain institutions or commercial entities to release data in a sanitized, structured way under an open license after a specific period would encourage wider access to data.

**Tech investments**

- One of the challenges in using openly licensed or public domain content as training data remains extracting it from PDF. Investing in better tools for extracting openly licensed content from difficult formats like PDFs and providing these tools as open-source software would accelerate access to quality training data for AI systems.
- The development of nuanced - as opposed to blanket - consent mechanisms for opt-outs before data collection, as well as for post-release removal could enable data right holders to distinguish between various uses of their data, potentially slowing the decline in open data availability.

**Open Questions:**

- Partnerships between LLM developers and cultural institutions remain rare. How can we strengthen collaboration to ensure data contributions to open LLM ecosystems, rather than exclusive agreements with large tech firms?
- Given the bias toward English content across LLMs and the difficulties to do justice to underrepresented communities, how can underrepresented communities participate in data collection, curation, and auditing, and what incentives would encourage data contributions or volunteer work?

**5.2. Clarity Over the Legal Status of the Data**

Legal uncertainty around data use—especially for volunteer-driven organizations without substantial legal support—remains a significant barrier and a chilling effect for the ecosystem.

**Policy interventions**

- A "safe harbor" provision across jurisdictions could help by allowing organizations to correct licensing errors without an immediate threat of legal consequences.

**Tech investments**

- Developing machine-readable standards for metadata across the Web would help clarify licensing and consent at scale, reducing legal risks for data users.

**Open Questions:**

- What tools and processes do we need to encode preferences in metadata so that they are machine readable, enabling legal certainty for dataset builders? Is it realistic to expect people to set such preferences on an individual level, or should the solutions also enable preferences setting and decision making at a group or community level or via content platforms and infrastructure providers who set preferences for their content?
- How should safe harbor protections be structured, and what would be an appropriate grace period for corrections?

### 5.3. Responsible AI governance

The openness of datasets alone doesn't guarantee positive societal impact or prevent possible harms. Responsible governance remains a crucial aspect of trustworthy AI.

**Tech and community investments**

- Non-English content is underrepresented in LLM training data. More builders should reach out to local communities to help build more high quality non-English data.
- Terms of Use are often difficult to understand, long, and not standardized. Moving forward, making them easier to interpret and machine readable, for example by creating standardized "modules" that can be combined to fit creators' needs, would make it easier to both respect, and enforce Terms of Use.

**Open questions**

- Removing harmful content from a dataset post-release is a big problem, as previously published versions often continue to be available online. How could the harm caused by older versions of datasets be minimized?
- How can we assess the environmental implications of open datasets, both in terms of potential [carbon emissions reduction](#) and carbon emissions expansion?

### 5.4. Sustainable funding

Open datasets are inherently made available without a cost which prevents their builders from relying on traditional business models. How can we make the ecosystem more financially resilient and sustainable without compromising on the open ethos?

**Policy interventions**

- To turn open LLM datasets and models into public goods, they ideally are also funded as such, at least in part to ensure long term sustainability.
- Policy makers could also help make open LLMs more competitive, for example by requiring openness of training data, model parameters and other elements for certain use cases or mandating public institutions to only use open LLMs.

**Open questions**

- What are viable routes to make open datasets sustainable long-term? Besides public funding, could income models like [Wikimedia Enterprise](Wikimedia Enterprise) or [Spawning's Source.Plus](Spawning's Source.Plus) provide needed revenue and rewards for right holders? Do we need to change the way we think about openness and introduce access limitations to prevent open data exploitation?
- Should tech companies contribute financially to cultural institutions that have digitized and preserved public domain data?

## 6. Conclusion

The production and maintenance of open datasets for LLMs is a complex, evolving challenge requiring active participation from diverse stakeholders. By creating common artifacts, best practices and standards, implementing thoughtful policies, investing in targeted technologies, and exploring sustainable funding models, we can foster an ecosystem where high-quality open data supports innovation while preserving the openness of the Web. Achieving this will demand collaborative effort and a commitment to the public good, but it is an essential pursuit to ensure that the next generation of AI remains transparent, accessible, and beneficial to all.

## Appendix A: Case Study EleutherAI's Common Pile

**Motivations and goals**

EleutherAI's primary interest is to improve transparency and interpretability of LLMs by creating standardized "default" datasets that are consistently reused for years. Using the same training data across models facilitates rigorous evaluations of their performance because it limits the number of factors that lead to variations between them. According to EleutherAI, Common Pile was developed mainly for two reasons:

1. Since the release of The Pile in 2020, EleutherAI gained significant experience in model training, including on how to better format data for it. As time passed, a bigger update of the first Pile became more desirable.

2. EleutherAI received feedback from a number of organizations that were interested in using The Pile, but were not able to for legal, ethical, or other reasons.[6] With public concern about copyright in AI training data mounting, EleutherAI concluded that a widely used, standardized default dataset needs to consist only of openly licensed content to ensure the widest possible adoption.

EleutherAI also views Common Pile as a statement against claims by some leading AI companies that training performant LLMs without copyrighted material is impossible. Such claims are self-serving to large companies who can shoulder the risk of legal uncertainty, as well as strike deals worth hundreds of millions for exclusive content licensing for their proprietary datasets. Meanwhile, smaller corporate actors, researchers, and public institutions depend on open datasets to compete. This way, Common Pile can contribute to making the LLM ecosystem more competitive and diverse.

**Data curation and processing**

Common Pile is made up of a series of subsets, similar to The Pile (which had 22 subsets), but these subsets are similar in size compared to its predecessor. Overall, Common Pile was curated to have a higher proportion of content that is known to correlate highly with model performance:

- Code: The Pile pioneered combining code and natural language in training data. Since then, research has shown that a higher proportion of code correlates with better performance, which is why Common Pile now includes an even larger code subset.

---

[6] EleutherAI openly stated that the dataset includes unlicensed data that was already in widespread use at the time it was published

- Public domain books: Creating a large subset of public domain books is one of the most challenging and time-consuming tasks in the development of Common Pile. Determining whether a book is in the public domain is a complex process for a single jurisdiction and becomes harder on an international level. As a US-based organization, EleutherAI therefore focuses on books whose copyrights have expired according to US copyright law, as is the case for all books published in the US before 1929. However, since historical texts can contain language, morals, and biases that may be deemed undesirable for LLM training, EleutherAI has also worked to identify and procure newer books which did not have their copyright renewed between 1929 and 1978. While a first large batch of public domain books will be included in Common Pile from the start, the identification and gathering of more books will take additional efforts to merge and reconcile bibliographic metadata from various library catalogs. The next big challenge is to gather the book texts in suitable formats. Many books are only available as PDFs and not all of those PDFs have high quality optical character recognition (OCR), so extracting the plain text at scale is technically challenging.

- Academic papers and other content with high information density: Similar to The Pile, academic papers were mostly sourced from preprint open repositories like arXiv or pubmed, and other collections of text with high information density, like StackExchange or Wikipedia.

- Government texts: Another important source of newer public domain content were texts produced by government bodies, since many are public domain by default by law.

- Common Crawl: Another big challenge in developing Common Pile was building a parser that goes through Common Crawl crawls and identifies license information. A lot of web pages mention their content licenses, but not in a standardized way that can easily be extracted from metadata. EleutherAI extracted Creative Commons licensed content that will be filtered further by adopting filtering pipelines developed in [Dolma](), [FineWeb](), or [MassiveText]() (yet to be decided at the time of writing). Sharing existing libraries and methodologies creates consistency and standards across organizations, which is in line with EleutherAI's goal to increase transparency and interpretability in the field.

**Dataset components**

Expanding on the categories above, this section details the design choices made in the process of producing the Common Pile on a per-component basis. While text processing and motivation for particular datasets is included in some cases, the primary focus is on licensing metadata and provenance.

The components range from still being obtained to being fully processed. For the work-in-progress components, we have sought to detail both what has been accomplished so far as well as what we plan on doing in the future. When available, data regarding the size of the component is provided.[7]

**Academic Texts**

Academic papers are a rich source of high-information density text about the world. Scientific research, in particular, is widely regarded as essential to training LLMs that are knowledgeable about the world. Due to differing cultural norms in social science and the humanities, it is challenging to build a large corpus out of research in those fields. Thus, the academic text here is overwhelmingly from physical and biological sciences, along with mathematics and legal documents.

*ArXiv*

We include papers from the ArXiv, a preprint server where research papers can be disseminated before peer review. The majority of papers published are categorized as Physics, Math, and Computer Science. All papers are uploaded by the authors, who are required to identify the license under which the paper is shared. We include papers that are licensed under the permitted licenses, which is about 15% of all papers. For each paper, the LaTeX source is converted to plain text, with math sections left as-is.

*BioDiversity Heritage Library*

The BioDiversity Heritage Library is the world's largest freely accessible repository of life sciences research maintained by the Smithsonian Institute. BHL contains texts with a wide variety of licensing, including in-copyright texts that are included under restrictive terms by agreement with the copyright owners. We include all documents whose metadata lists them as being under an acceptable license. BHL also provides a due diligence statement for some texts: "No known copyright restrictions as determined by scanning institution." We do not currently include such texts pending further investigation into the extent of the due diligence checks performed in such cases.

*Case Law Access Project*

Ravel Law, Harvard Law Library, and the Harvard Library Innovation Lab digitized over 40 million pages of U.S. court decisions consisting of 6.7 million cases from the last 360 years of American case law into a dataset that is widely accessible to download. The scanned pages were OCRed from 40,000 physical volumes with ABBYY FineReader PDF. Metadata such as citations and dockets were additionally revised by human reviewers. Accessing a bulk download of the raw data can be done through the [Case Law Access Project API (CAPAPI)](#).

---

[7] Completed components can be found on the Hugging Face Hub. All processing code can be found at https://github.com/r-three/common-pile.

During the digitization of these scans, there were OCR errors that occurred. Each of the extracted texts was post-processed to fix encoding, normalization, repetition, redundancy, parsing, and formatting errors. A post-processed version of the Case Law Access Project data with corrections is available on the [Hugging Face Hub](#). The content of the Case Law Access Project is licensed CC0 or is in the public domain.

*CourtListener*

CourtListener is a bulk legal domain database consisting of financial disclosures, judge information, oral arguments, U.S. case law, and opinions, developed and maintained by the nonprofit organization, Free Law Project. Opinions contain both concurring and dissenting texts from court judgments that encompass reasoning, and principal or majority opinion. The opinion data is collected from numerous publicly available sources such as the U.S. Supreme Court, U.S. State Supreme Courts, U.S. District and Bankruptcy Court, and U.S. Court of Appeals. The financial disclosure data contains 32,336 documents that consist of information regarding 1,901,720 investments. The raw data can be bulk downloaded [via an API](#).

Additional post-processing was done to properly extract, normalize, and fix texts, correct OCR errors, remove boilerplate and repeating characters, and fix tags. The content of CourtListener is licensed CC0 or is in the public domain.

*PubMed Central*

PubMed Central is an open-access archive of biomedical and life sciences research paper maintained by the U.S. National Institutes of Health's National Library of Medicine. Similar to the BHL, PMC contains texts with a wide variety of licensing, including in-copyright texts that are included under restrictive terms by agreement with the copyright owners. Unlike BHL, articles and article metadata in PubMed are sourced exclusively from the publishing journals.

**Code Data**

One of the most widely used applications of large language models today is for code generation and analysis. Ever since [Gao et al. (2020)](#), it has been common to train models on code and natural language simultaneously. It is sometimes claimed that this promotes logical reasoning in non-code domains, though we think the evidence of this is ambiguous. Regardless, it's clear that training on code data promotes widely desired use cases.

*Stack v2*

The [Stack v2](#) is a dataset of GitHub code spanning 619 programming languages created as part of the [BigCode project](#) led by Hugging Face and ServiceNow and in collaboration with Software Heritage. We take the variant that they used for the training of StarCoder2 and filter it for our desired licenses. For further information on the license identification and metadata protection procedures, see [Lozhkov et al. (2024)](#).

*Stack Exchange*

We include data from the [Stack Exchange data dump](#). Content posted on Stack Exchange sites is licensed under [CC BY-SA](#); the specific version has changed over time. The question-and-answer format of Stack Exchange sites dovetails nicely with common use-cases of LLMs. We also include the "meta" version of each Stack Exchange site. These sites include meta-level questions about the management of the associated Stack Exchange site. We include these as they tend to be more conversational. Each document in the datasets consists of a "Question," followed by each of the "Answers," sorted according to their votes at the time of the dump. If one of the answers is marked as the "accepted answer," it is always the first answer to appear. Comments on the question and answers are included, but can be removed during the data generation process.

**Public Domain Books**

Books are an essential source of long context, semantically complex, text. Most LLM trainers view large books corpora as an essential source of training data. However, this poses a massive challenge from a licensing POV as the overwhelming majority of books are not in the public domain. Additionally, there's a strong desire for more recent texts as historical texts can reflect cultural contexts vastly different from today's world.

*Project Gutenberg*

[Project Gutenberg](#) includes digitized versions of many books from western literature. We include books whose metadata indicate they are in the public domain, as well as the books from the PG19 dataset—a collection of books published before 1919 and have thus aged into the public domain. The length of the included books should be helpful when pre-training due to the existence of long-distance context.

*Library of Congress Books*

This component contains more than 130,000 English books digitized by the Library of Congress (LoC) as part of the [Selected Digitized Books](#) collection. All works from this collection were determined by the LoC as being free of known copyrights and belonging to the public domain in the United States. The books were retrieved through the LoC API.

*Pre-1929 Books*

All books published in 1929 or earlier in the United States are in the public domain today. While some of these works are contained in Project Gutenberg, there's a substantial amount of this work that is not. We have been processing books available from a variety of repositories that are listed as being published before 1929 in the United States and are verifying their metadata with bibliographic data from various libraries.

*Post-1929 Unrenewed Books*

For works published between 1929 and 1964, copyright owners were required to renew their copyrights after 20 years. Any works that were not renewed are now in the public domain. Unfortunately, there is no high-quality record of which works were renewed, making most of these works in practice not usable. The New York Public Library engaged in an excellent project to digitize and release copyright and copyright renewal records, providing public access to the requisite information to figure out which books between these dates are in the public domain.

We are building on the NYPL's work to create a comprehensive list of public domain books. The data released by the NYPL includes two key items: a list of copyright registrations and a list of copyright renewals. Our ultimate goal is to match these entries to determine which books were never renewed. This task requires substantial effort due to issues such as OCR quality, the lack of a consistent ID system used by the copyright office, and incorrectly filled-out forms. Our current pipeline is shown below in Figure 3.

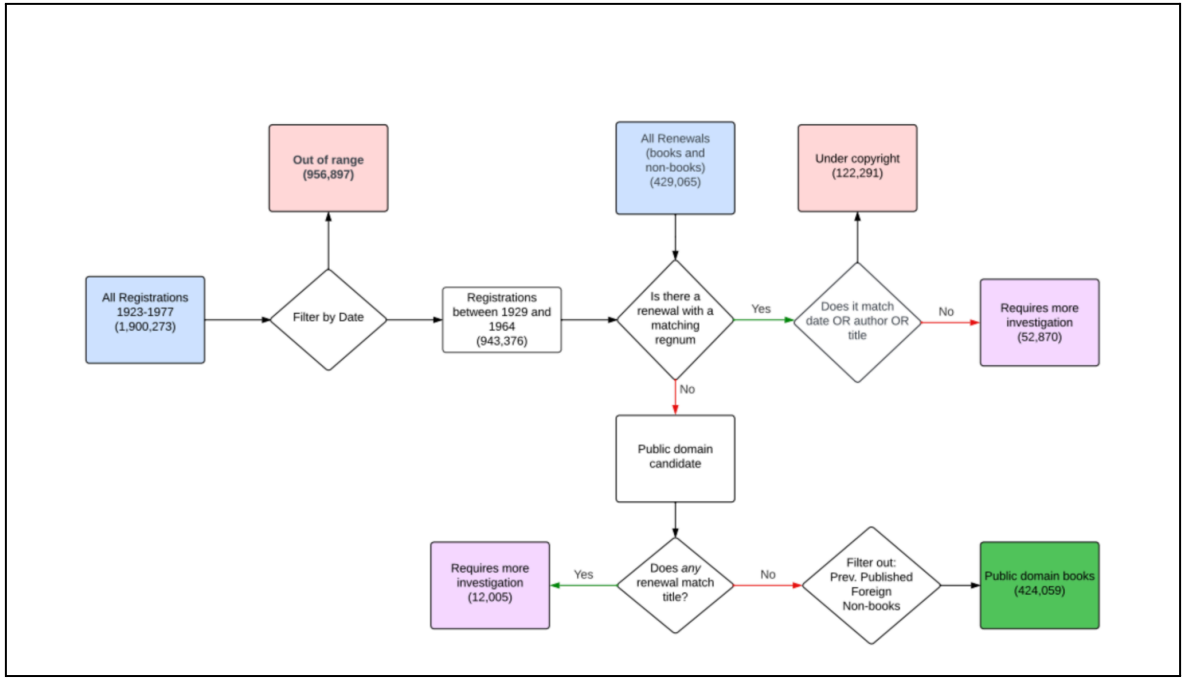

*Figure 3: Flowchart illustrating the process of analyzing copyright office records to identify books in the public domain due to lacking a renewal. The two input nodes are labeled in blue and the output nodes are classified as being clear exclusions (red), requires more investigation (purple), or believed to be in the public domain (green). Each node is labeled with the number of works at that stage.*

Some renewals contain copyright metadata in an unstructured format. We use a large language model trained by Nous Research to process this unstructured data and extract the relevant copyright metadata fields. To validate the model's accuracy, we test it on works for which we have metadata in a structured format. The large language model demonstrates high accuracy in extracting key copyright metadata from unstructured data, achieving an impressive 96.58% accuracy for titles and 97.20% for registration numbers. The accuracy for authors (90.25%) and dates (87.66%) is slightly lower, but mistakes in these values cannot wrongly cause a work to be classified as in the public domain. Assuming that errors in identifying titles and registration numbers are independent, the odds of a work being wrongly classified in the public domain due to incorrect LLM-identified metadata is 0.09%, or about 380 books out of 424,059.

We are also experimenting with using a language model to assist with investigating the purple "requires investigation" boxes. Preliminary experiments indicate that the model has an accuracy of 98.6% in determining if an ambiguous match is correct or not. Once we have a list of books that we believe are in the public domain, we then need to find their full text. The copyright office metadata does not reliably have useful ID numbers, so we are utilizing a range of resources to obtain cross-reference ID numbers and identify libraries or other sources that possess these texts. We are currently working on obtaining full-text copies of all books we believe to be in the public domain. We have identified approximately 500,000 books that we believe to be in the public domain, though it is currently unclear how many have accessible digitizations.

**News**

News-wire text has a long history of use in NLP research but often comes with restrictive licenses. Due to the quality of news-wire text and the focus on real-world events and facts, we include permissively licensed news-wire text.

**Web Data**

While we have primarily aimed at identifying high-quality text from trusted sources, supplementing that data with more generic web data ensures that we have a comprehensive dataset.

*Common Crawl*

Website publishers may choose to license the content under one of the Creative Commons licenses or host content that is in the public domain. Using 52 crawls provided by Common Crawl, we curate a set of high-quality web pages that match these licenses. We limit this subset to Creative Commons licenses as we found automatically detecting all Blue Oak Council Bronze or higher licenses to be infeasible due to the wide variety of ways people describe their licensing terms. By limiting to Creative Commons licensing, we can rely heavily on regex matching for identifying licenses. Creative Commons pages are typically labeled using a standardized identifier placed in the HTML code of [a website](). Similarly to [Habernal et al.](), we use a regular expression to identify these elements. Then, the content of each page is extracted into plain text using the [Resiliparse library](). We are currently in the process of carrying out a detailed study of the end-to-end accuracy of our license identification pipeline. After initial filtering, we further refine these subsets by applying heuristics that are commonly employed when curating text for language model pre-training to maximize the quality of the obtained text. After these filtering steps, we end with 259,728,610 pages comprising 221,715,271,483 words.

*Data Provenance Collection*

There are thousands of carefully curated datasets designed for training specialized NLP models. They are highly heterogeneous in their sources, creators, and purpose, as well as their licensing conditions. Using the [Data Provenance Explorer,]() we filter datasets based on a series of conditions: contains English or code data, is not model generated, and has a permissive license. The license is identified as the one on the dataset's GitHub page (if available), or the license identified by the Data Provenance Initiative, where

33

annotators searched over the dataset's associated academic paper, GitHub, Hugging Face, and website pages for a dataset-specific license.

*Foodista*

We include data from [Foodista](#) as it is licensed under CC BY/3.0. It includes blog articles about food; informational pages on ingredients, tools, and techniques; and recipes. Recipe generation is a common use-case of current LLMs, and the step-by-step nature is akin to the formatting used in instruction tuning.

*Public Domain Review*

The [Public Domain Review](#) is a collection of long-form essays about works that have recently entered the public domain. They are licensed under [CC BY-SA/3.0](#).

*Ubuntu IRC*

The chat logs of IRC channels about the Ubuntu operating system on the Freenode [IRC chat server](#) are released into the public domain. We include this data as we expect that human-to-human dialogue will be useful for pre-training as many LLMs are used in consumer-facing chat applications. Each document in the dataset is the chat log from a single channel over the course of a day. We filter out messages from bots and channel announcements, such as when a user changes their name.

*Wikis*

Wikis are community-managed knowledge bases, generally about a specific topic, created via collaborative edits from community members. The most common software used for creating a Wiki is MediaWiki, and wikis are published with permissive licenses. Often many wikis created by diverse communities are hosted in one place by a third party. These are termed "Wikifarms" and make it possible to create a very large corpus of text from wikis.

We include wikis published by WikiMedia—Wikipedia, Wiktionary, etc.—which regularly publish XML dumps of their sites. Additionally, The WikiTeam project conducted by ArchiveTeam has been consolidating and archiving wikis from the official WikiMedia Project, ones produced using the MediaWiki library, as well as from Wikifarms in an Internet Archive collection.

We also include the "talk" pages from each wiki. These are meta-pages where editors discuss changes to the main content of the wiki. We hope that the conversational nature will help LLMs trained on this data when used in chat applications.

*YouTube Transcripts*

There is an immense amount of conversational data available on YouTube. We are currently searching YouTube to identify videos based on uploader-specified licenses and then running the identified videos through a Whisper-based pipeline to obtain high-quality text transcripts. We have not yet validated the correctness of the uploader-specified licenses. This dataset is similar to the recently released [Pleias YouTube Commons dataset](#) but with additional licenses beyond CC-BY and higher quality transcription provided by Whisper instead of the default YouTube auto-transcription.

**United States Government Documents**

All U.S. Government-authored or produced data and information are considered to be in the U.S. Public Domain. This represents an immense and under-leveraged source of public domain text.

*Patent Office*

The United States Patent and Trademark Office (USPTO) has received approximately 20 million patent applications. The dataset encompasses the totality of these applications, making it a rich resource for training language models. The structured nature of patent documents, which adhere to a very specific format, provides valuable examples for understanding technical language, legal terminology, and formal documentation styles. Patent applications are particularly valuable for training models due to their long-context dependencies, where the claims must follow detailed descriptions and prior references within the document.

*Government Publishing Office*

The U.S. Government Publishing Office is the agency responsible for publishing documents authored by the U.S. federal government. All documents released by the GPO are in the public domain.

**Data governance**

Because Common Pile is meant to become a default dataset that should be reused for years to allow for better model comparison, there are currently no plans to update and change it after publication. Moreover, whatever license the content was originally published under applies to the subsets. EleutherAI is not adding any terms of use.

## Appendix B: Case Study: Pleias' Common Corpus and YouTube-Commons

**Motivations and goals**

Similar to EleutherAI, Pleias views the publication of Common Corpus and YouTube-Commons as a statement to show that there is plenty of openly licensed content available that just has to be made consumable for LLM training. To unlock such content, Pleias hopes that their work on Common Corpus and YouTube-Commons can contribute to a "data commons" for AI that is powered by a community that continuously improves and expands the datasets it hosts, similar to how Wikipedia's community does. As Pleias' co-founder Pierre-Carl Langlais said in our interview, the status quo around generative AI is detrimental to a digital commons, as there are so many incentives to close down access to datasets and enter into exclusive data licensing deals, or to restrict or limit how the content can be used for AI training. Moreover, leading AI companies might benefit disproportionately from open datasets due to [LLM scaling laws](#), which is why Pleias is simultaneously pushing for smaller models trained on fully open datasets to be more competitive.

A key assumption at Pleias is that there is a market for such smaller models, especially for the public sector. The public sector in France, where Pleias operates, has requirements for the transparency of the technology it uses to process data about citizens. There is also interest in increasing local technological sovereignty, which makes relying on APIs from leading U.S.-based AI companies unattractive compared to using smaller models built locally with open source components.

More widespread adoption of smaller models could help attract volunteer contributors to AI data collection and curation, which could then help to unlock more freely available content for LLM training over time. Long term, the hope is that it would also make training data more linguistically diverse because of the more localized, community-driven contributions worldwide.

**Data curation and processing**

Different from the books subset in EleutherAI's Common Pile, Pleias deliberately only includes older public domain content in its first release of Common Corpus. Pleias faced an issue similar to EleutherAI: Clarifying whether content was under public domain across jurisdictions was time-consuming and required case-by-case investigation, as different regulations for when content enters the public domain apply in the U.S. and Europe, the two regions they focused on for the initial release. The company mostly limited itself to content published before 1884 as a precaution. To identify repositories of public domain content, it relied on a network of collaborators in the U.S. and across Europe, including the [French Ministry of Culture](#), [Occiglot](#), [EleutherAI](#), [HuggingFace](#), and [Nomic AI](#). The process of compiling Common

Corpus was iterative. It started with the French subsets only and slowly expanded to subsets in [other languages](#) created with local partners. Beyond identifying public domain content, a huge challenge (that EleutherAI faced as well) was turning the content into plain text for LLM training. Most documents were only available as PDFs. Pleias limited its data collection to PDFs with pre-existing OCR, but OCR quality was often low. Reliably extracting high-quality plain text from low-quality OCR is an ongoing challenge, and a focus of Pleias' work is on [post-OCR correction](#) to further improve the quality of Common Corpus and similar datasets.

YouTube-Commons is meant to complement Common Corpus, which contains many formal texts, with conversational data. The process was relatively straightforward: Realizing that YouTube contains a huge amount of videos under Creative Commons' CC-BY license, Pleias created a dataset consisting of the video transcripts (including automatic transcripts and translations) and metadata provided by YouTube. While this approach is not considered ideal because the quality of the transcripts and especially the automated translations varies, it is viewed as a first step that a community could help improve and expand upon. The metadata about the videos was included for licensing reasons and to make YouTube-Commons useful for creating openly licensed multimodal training datasets in the future.

**Data governance**

In stark contrast to EleutherAI's approach of creating a stable default dataset, Pleias aims to foster a community that constantly improves and expands Common Corpus and YouTube-Commons following a "release often, release early" mindset.

This imagined community does not yet exist, and most of the work is still done by only a handful of people. However, Pleias hopes that eventually, a community similar to those in the open science and open data space will emerge, with different stakeholders (both institutional and non-institutional) bringing stability and expertise, and ensuring continuity. Pleias expects to attract such stakeholders by demonstrating that the data commons it hopes to create is mutually beneficial. At the moment, some cultural heritage organizations are cautious about Common Corpus because it repurposes content they digitized in ways they do not control. Pleias hopes that its work on post-OCR corrections will help shift mindsets because bad OCR is also a huge issue for cultural heritage projects that want to enable large scale analysis of the content they host. The idea is to focus on such overlapping interests between AI users and content curators.

Like EleutherAI, Pleias does not impose any terms of use on its datasets and whatever license the content was originally published under applies. There was also no further curation of the contents of the data, meaning that no YouTube videos under CC-BY license were excluded due to their contents, for example.